\documentclass[twocolumn,showpacs,preprintnumbers,amsmath,amssymb]{revtex4}
\usepackage{graphicx}
\usepackage{dcolumn}
\usepackage{bm}

\begin{document}

\title{Varying Cu-Ti hybridization near the Fermi energy in Cu$_{x}$TiSe$_{2}$: Results from 
supercell calculations.}

\author{T. Jeong and T. Jarlborg}

\affiliation{
DPMC, University of Geneva, 24 Quai Ernest-Ansermet, CH-1211 Geneva 4,
Switzerland
}


\begin{abstract}
The properties of Cu$_{x}$TiSe$_{2}$ are studied by band
structure calculation based on the density functional theory for 
supercells. The density-of-states (DOS) for $x$=0 has a sharply raising
shoulder in the neighborhood of the Fermi energy, $E_F$, which can be
favorable for spacial charge modulations. 
The Cu impurity adds electrons and
brings the DOS shoulder below $E_F$. Hybridization makes the Ti-d 
DOS at $E_F$, the electron-phonon coupling and the Stoner factor very large.
Strong pressure dependent
properties are predicted from the calculations, since the DOS shoulder
is pushed to higher energy at a reduced
lattice constant. Effects of disorder are also expected to be important
because of the rapidly varying DOS near $E_F$.

\end{abstract}


\maketitle


The layered TiSe$_{2}$ compound is much studied because of the appearance of
charge density waves (CDW). The ground state is believed to be either a 
semimetal or a semiconductor 
with a small indirect gap\cite{freeman,zung,traum}. From photoemission experiments 
it is concluded that the CDW transition consists of a change in electronic 
structure 
from a small indirect gap 
into a state with a larger indirect gap at a 
slightly different location in the Brillouin zone\cite{kidd}.
Recently Morosan {\it et al.}\cite{morosan} reported that a continuous 
intercalation of TiSe$_{2}$ with Cu is possible, where the Cu atoms
enter between the TiSe$_2$ layers to 
yield Cu$_{x}$TiSe$_{2}$, with solubility of Cu up to $x \approx 0.11$.
 The CDW transition is continuously 
suppressed, and a superconducting state appears near $x=0.04$ 
with a maximum T$_{c}$ of 4.15 K at $x=0.08$. 
A CDW superconductivity phase diagram as a function 
of doping is developed for Cu$_{x}$TiSe$_{2}$, analogous to 
the antiferromagnetism-superconductivity phase diagram found 
for the high temperature superconductors.

In this work
we investigate the electronic structure of 
Cu$_{x}$TiSe$_{2}$ for $x$=0 and $x \approx 0.08$ based on the density functional theory  
applied to supercell calculations.
As will be shown, one of the main results is that Cu doping introduces electronic
charge carriers, leading to high-DOS properties, which partly confirms the interpretation
of experimental results \cite{morosan}.

\begin{figure}
\vskip -5mm
\includegraphics[height=8.5cm,width=8.5cm,angle=-90]{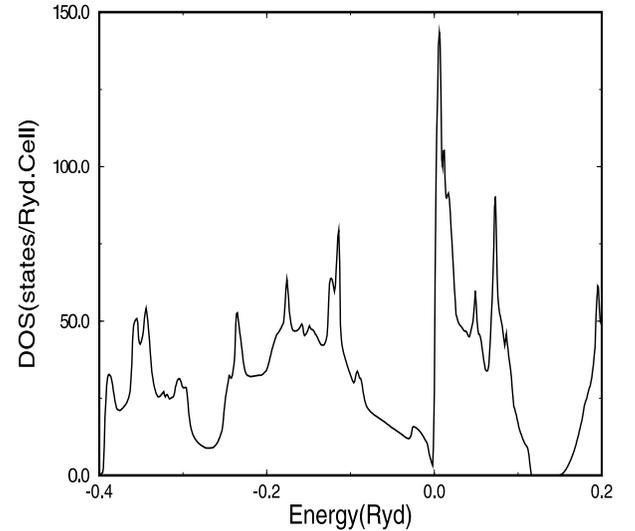}
\caption{The total density of states of 
TiSe$_{2}$. The energy is relative to $E_F$.}
\label{fullband}
\end{figure}

\begin{figure}
\vskip -5mm
\includegraphics[height=8.5cm,width=8.5cm,angle=-90]{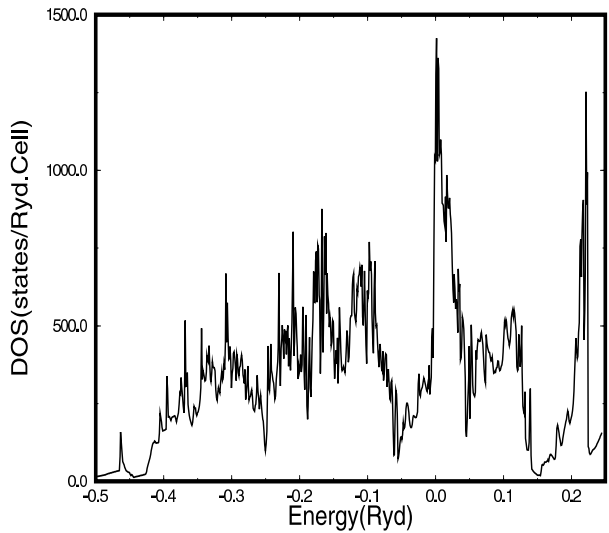}
\caption{The total density of states of CuTi$_{12}$Se$_{24}$ 
at the lattice constant 6.7 a.u.. The energy is relative to $E_F$.}
\label{dos3}
\end{figure}

\begin{figure}
\vskip -5mm
\includegraphics[height=8.5cm,width=8.5cm,angle=-90]{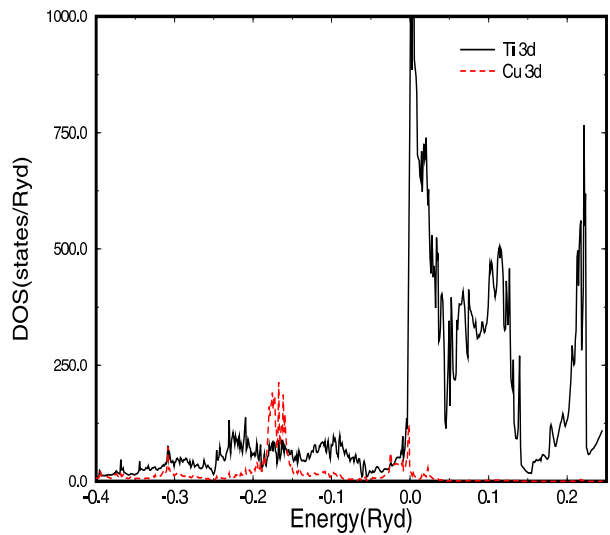}
\caption{The projected Ti-3d and Cu-3d parts of the total DOS 
shown in fig. 2. 
 }
\label{dos}
\end{figure}
 
\begin{figure}
\vskip -5mm
\includegraphics[height=8.5cm,width=8.5cm,angle=-90]{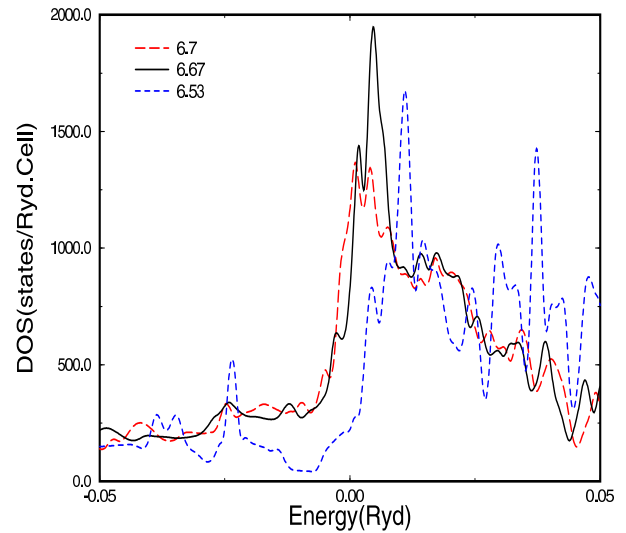}
\caption{The total density of states of CuTi$_{12}$Se$_{24}$ near $E_F$ (at zero energy
in each case)
at three different lattice constants
. }
\label{dos4}
\end{figure}


TiSe$_{2}$ has the layered CdI$_{2}$ crystal structure. 
Planes of titanium atoms are surrounded by planes of selenium 
atoms, forming a sandwich structure,
where the selenium atoms are arranged in octahedral coordination
around the titanium atoms at $z=\pm \frac{1}{4}$.
The unit cells are arranged so that the metal atoms are 
stacked directly above each other.  
The TiSe$_2$ layers are well separated
and bound by van der Waals interaction.
As Cu atoms are added, they occupy positions within
these open spaces between 
the TiSe$_{2}$ layers. 
This results in a systematic expansion of the unit cell 
with increasing Cu content in Cu$_{x}$TiSe$_{2}$.
The experimental lattice constants are a= 6.7 $a.u.$
and c=11.35 $a.u.$ when $x$ is near 0.08 \cite{morosan}.


The self-consistent Linear Muffin-Tin Orbital 
method \cite{lmto,bdj} is applied to a 48-site supercell in which one Cu atom is
intercalated between two Se layers. In these calculations we
use a potential based on the local density approximation (LDA) \cite{lda},
with no special on-site correction due to correlation. The cell consists of
CuTi$_{12}$Se$_{24}$  and 11 empty spheres in the
open space separating the TiSe$_2$ layers. The repeat distance
between Cu atoms is 2 unit cells in the two plane directions
and 3 cells along $\hat{z}$. This is chosen to minimize the
Cu-to-Cu interactions within a reasonably simple unit cell for a
Cu concentration $x$ of 0.0833, which is close to the concentration
for optimal $T_C$.
The self-consistent iterations use initially 8 $k$-points 
in the irreducible Brillouin Zone, while they are 
terminated by iterations using 75 $k$-points.
The Wigner-Seitz sphere 
radii are 2.88 a.u. for Ti and 3.17 a.u. for Se respectively.
The basis-set for the atoms includes d-states in the direct basis and f states
as tail contribution. The direct basis for empty spheres includes p, with d states
in the tails.
The bands are calculated at the experimental lattice constant, and
at two smaller volumes, for
the case with optimal Cu insertion. It should be recalled that
LDA typically underestimates the lattice constant by 1-3 
percent for 3d metals.

The supercell is repeated periodically and effects of random disorder,
of distortion around the impurity, or of higher order periodicities
among the Cu atoms are not included. It should be noted that any
disorder (structural or substitutional) within the supercell will
introduce band broadening in general, since degenerate bands of
the perfectly repeated supercell will be lifted. 
These effects of disorder
are likely to be more important for this particular
system, because of the very sharp DOS structures near $E_F$.

The electron-phonon coupling $\lambda$ is calculated using the
Rigid-Ion approximation and experimental information of the 
phonon moment $\langle \omega^2 \rangle$. As the Debye temperature for 
TiSe$_2$ is not known to us, we use an average
Debye temperatures of NbSe$_2$ (280 K \cite{nbse}) and
TiS$_2$ (235 K \cite{tis}). The modification of the pure
phonon contribution to $\lambda$ due to the additions of
Cu is probably not very important. Since the DOS is large
on Ti, it is expected that the electronic contribution will
be dominated by these sites. The pressure variation of
the phonon moment is scaled through the calculated variation
of the Bulk modulus. These procedures are not very precise,
but they should be sufficient for getting insight of the contributions to 
$\lambda$ \cite{pjp}. 

The Stoner factor, $\bar{S}$, is calculated from the paramagnetic
band results in the standard way, 
by assuming a finite rigid-band splitting due to exchange of all bands
\cite{jf}. If there is only one band contributing to $N(E_F)$, one can write
$\bar{S} = N \cdot I$, where $I$ is an exchange integral for that
band. The DOS is dominated by the Ti-d band and the Stoner factor
is mainly because of Ti-d electrons. Hybridization with other bands, and the
induced exchange splitting on other sites, is likely to
reduce the effective Stoner enhancement, but spin-polarized calculations would be needed
to verify this. 


\begin{table}[b]
\caption{\label{table1}The calculated parameters of CuTi$_{12}$Se$_{24}$. 
The Debye temperature $\theta_{D}$ is from experiment and scaled through the calculated
volume dependence. The units of $\gamma$ is (mJ/mol K$^2$),
and N(E$_{F}$) is in units of states per (Ryd$\cdot$cell).}
\vskip 5mm
\begin{center}
\begin{tabular}{|l|c|c|c|c|c|c|}
\hline
$a_0$ (a.u.) & $\lambda$ & T$_c$(K) & $\bar{S}$ & 
$\gamma$ & $\theta_{D}(K)$ & N(E$_{F}$)\\
\hline \hline
6.7  &  1.52 & 20.2  & 1.31 & 11.13 & 240 & 1220\\
6.67 &  0.37 & 0.11  & 1.09 & 4.75 & 250 & 960\\
6.53 &  0.21 & 0.01  & 0.49 & 1.26 & 290 & 290\\
\hline
\end{tabular}
\end{center}
\end{table}

\begin{table}[b]
\caption{\label{table2} D-DOS at $E_F$ on Cu and Ti near (Ti$_{Cu}$) and far from (Ti) the
the Cu impurity. The units of N(E$_{F}$) are 
states per (Ryd$\cdot$atom) }
\vskip 5mm
\begin{center}
\begin{tabular}{|l|c|c|c|}
\hline
$a_0$ (a.u.) & Cu $d$ N(E$_{F}$)& Ti$_{Cu}$-$d$ N(E$_{F}$)& Ti-$d$ N(E$_{F}$)  \\
\hline \hline
6.7  &  22.1 & 127.  & 63.0 \\
6.67 &  19. & 72.   & 41.  \\
6.53 &  0.85 & 28.1  & 1.0 \\
\hline
\end{tabular}
\end{center}
\end{table}

The DOS for the primitive cell of undoped TiSe$_2$ is shown in Fig. 1.
An independent calculation using the Full Potential Local Orbital (FPLO) \cite{koepernik} 
method, using a potential based
on the gradient corrected
density functional \cite{pw}, 
gives a similar band structure, but with a slightly larger dispersion
of a band above $E_F$, so that the first unoccupied DOS peak becomes less prominent.
The band results indicate that the pure material is a
semimetal where $E_F$ is in a valley in the DOS, with a low value of
$N(E_F)$ of 10-20 states per Ryd per TiSe$_2$. The uncertainty is due to the
fact that $E_F$ falls slightly on the right hand side of the minimum of the DOS (see Fig. 1), 
on the steep shoulder, where the DOS increases rapidly
with energy. This is for the perfect structure without disorder or CDW.

A simple rigid-band model for a potential (or charge) modulation in a material
with an increasing DOS near $E_F$ (as for pure TiSe$_2$)
can be shown to favor CDW. Shifts of the local DOS functions relative to $E_F$
suggest that some regions of the material should get a high local N$(E_F)$ 
because of the large DOS just above $E_F$. Oppositely, other regions where the potential
is repulsive will have $E_F$ closer to the local "gap" with minimum DOS. A similar reasoning
as for a Peierls transition show that the total kinetic energy will be lowered by
such a splitting of local states. Namely, since the DOS is sharply raising at $E_F$
and the number of states is conserved we find that an upward shift, $\delta_u$, of the
DOS relative to $E_F$ within the regions that loose electrons will be larger than
the downward shift, $\delta_d$, of the DOS within the region that gain electrons.
If the average DOS to the left and right side are $N_u$ and $N_d$, respectively, and $\Delta n$
is the number of transfered electrons, we have that $\Delta n = \delta_u \cdot N_u = \delta_d \cdot N_d$.
But the kinetic energy will be 
$\frac{1}{2} \Delta n (\delta_d - \delta_u)$, i.e., negative. Or
equivalently, the total (kinetic) energy will be lower for the CDW state.
If, by electron doping, $E_F$ moves to higher energy on a linearly increasing DOS,
then the relative difference between $N_u$ and $N_d$ will be smaller, and the gain
in kinetic energy and the tendency for CDW formation will diminish.
 
Total energy contributions from the
Coulomb interactions and hybridization are neglected here, but the model
gives a hint that a sufficiently raising DOS near $E_F$ is favorable for CDW. Nothing
can be said about the periodicity of these waves from the model. However, from qualitative
arguments it is expected that modulations
with long wave lengths are more probable, since they have lower costs in terms of hybridization energy.
It can also be noted that irregular regions of up- and down-shifted potentials,
as for long-range disorder,
will be favored through the same type of energy gains as for CDW. 
Supercell calculations with some guidance for
probable CDW modulations would be needed to solve this problem.
It is generally believed that Fermi surface
nesting is not a determining factor for the k-vector of the CDW \cite{kidd}.

When Cu is inserted it acts as an electron dopant. Cu 
brings along d-states near $E_F$, which hybridize
strongly with the Ti-d band, and the total DOS at $E_F$ increases very
much as can be seen in Fig. 2. From an inspection of the local Ti DOS
functions it can be verified that the Ti close to the Cu site has the
highest $N(E_F)$ values. The total Cu-d and the sum
of all Ti-d DOS functions are shown in Fig. 3.

The DOS peak near $E_F$ is high and narrow. This makes 
any calculation of $N(E_F)$ dependent properties very sensitive
to small shifts of $E_F$ (non-stoichometry) and to band broadening
(disorder effects). The energy derivative of $N(E_F)$ is high,
by shifting $E_F$ 2 mRy to lower energies, $N$ can be reduced
by a factor of two.

Nevertheless, by using the unshifted unbroadened band results
we obtain a very large Stoner enhancement, $S=1/(1-\bar{S})$, see table I. 
Ferromagnetism is expected when
$\bar{S}$ is larger than one. Apart from non-stoichometry and disorder effects
it is possible that the impurity-like electronic structure with a large
DOS near the Cu site will overestimate the calculation of $\bar{S}$. The 
reason
is that the exchange splitting is assumed to be equal on all sites in the
model, while in reality this may cost additional kinetic energy on sites
with low DOS. A spin-polarized calculation would give a better answer, but it
has not been attempted at this stage because of the very slow convergence for
the supercell.

The calculated values of $\lambda$ are reasonable and not as extreme as could
be expected from the large DOS and $\bar{S}$. The main contribution to 
$\lambda$
comes from the Ti atoms. The Ti-DOS is large, but there is only a moderate 
dipole
scattering contribution to the electronic part of $\lambda$, because the
total DOS is dominated  by the d-character (Typically 98 percent of the Ti-DOS is
of d-character).

A surprising effect is the large variation of $N(E_F)$ with pressure, $P$.
Two calculations at smaller lattice constants show clearly
that the narrow peak of the Cu-d state moves sensitively with volume.
This state is hybridized with the Ti-d band so that also the Ti-d DOS
at $E_F$ will vary much with volume. A downward shift of this peak by 
about 5 mRy is found when the lattice constant is increased from 6.67 to 6.7
a.u.. If the lattice constant is reduced to 6.53 a.u. there will be an
upward shift of the DOS edge to a location above $E_F$ and the total $N(E_F)$
is not much larger than for pure TiSe$_2$. The charges on Cu and Ti
increases with $P$ at the expense of charge on Se. The effect is that the Cu-d
band becomes more filled for a uniform reduction of volume. The peak in the
DOS 20 mRy below $E_F$ for the smallest lattice constant, is mainly a Cu-d state.
The Ti-d DOS is too far away in energy to hybridize with this state,
and at $E_F$ there will be less Cu-d to Ti-d hybridization.
This will result in a reduction
of $N(E_F)$. An inspection of the local Ti-DOS
for the smallest lattice constant reveals an important proximity effect near
the Cu site. The Ti-d DOS at $E_F$ 
is of the order 28 states/Ry/atom on each of
the 8 Ti on the layers closest to Cu. But in the more distant Ti layer,
where there are 4 Ti atoms, the local Ti-d DOS at $E_F$ is only of the order
1-2 states/Ry per atom, which is like the case for pure TiSe$_2$. The
reason is that the local Ti-d DOS is pushed just above $E_F$ for these sites. 
In the calculations for the two larger lattice constants the local Ti-d
DOS values at $E_F$ are comparable between all Ti sites, also within the most
distant layer. This shows that the properties of Cu doped TiSe$_2$ will
be very sensitive because of the exact position of the DOS shoulder
relative to $E_F$.

These calculations are made for perfect scaling of the supercell lattice.
Thus no effects of possible rearrangement of the Cu occupations, different
c/a ratio or disorder are taken into account.
If such effects are small, one can expect strong $P$-variations
of the properties. According to our calculations the system would
pass from ferromagnetism, towards exchange enhanced paramagnetism,
through a region with superconductivity and large electron phonon
coupling, and finally to a fairly low-DOS material when the volume
is reduced by about 8 percent only. However, this unusual sensitivity
is caused by a sharp DOS shoulder which is passing through $E_F$, and
therefore one can also expect large error bars. As mentioned before,
disorder and different Cu impurity distributions can be important and
have a moderating effect on the $P$-dependence.

 More hybridization with the Ti-$p$ 
and $f$
bands is needed for a larger $\lambda$. By reducing the lattice constant
it is possible to increase the hybridization and the electronic contribution
to $\lambda$, but the effect is masked by the changes in total $N(E_F)$,
which is decreasing much at large pressure. In addition, fine structures in 
the
peaked DOS add some noise to the evolution of $N$ with pressure.

An estimation of the superconducting $T_C$
from the calculated electron-phonon coupling via the McMillan equation gives
reasonable values in line with observations, see Table I. However, some reduction is 
expected
from spin-fluctuations, especially since the calculations indicate large 
exchange
enhancement. The calculated electronic specific heat constant, $\gamma$, includes
the enhancement for electron-phonon coupling, but not from possible
spin fluctuations. At the largest lattice constant $\gamma$ is considerably
larger than the measured value, about 4.5 mJ/mol K$^2$ \cite{morosan}. The calculated
result for $\gamma$ at the intermediate volume agrees best with experiment,
but the electron-phonon coupling is rather low at the same volume.
Again, this is an indication that
disorder in the real material will reduce the effective DOS at $E_F$
and increase hybridization.

In conclusion, we find that
undoped TiSe$_2$ is a low-DOS material with semi-metallic properties,
but the DOS is rapidly increasing near $E_F$. 
The latter fact is probably important for the properies of pure and 
doped TiSe$_2$. For instance,
the kinetic contribution to the total energy will be lowered by
spacial modulations of the potential as produced by a CDW or disorder. Electron doping,
accompanied by disorder, will put $E_F$ closer to the broadened DOS peak.
This should weaken the mechanism for CDW formation,
while the material soon becomes a high-DOS material. 
The local Ti-d DOS at $E_F$ for a supercell of Cu doped TiSe$_2$  
is increased very much through hybridization with the Cu-impurity and its 
d-band. The exact value of $N(E_F)$ is uncertain because of the sharp structure
of the DOS near $E_F$, which is found both for pure TiSe$_2$ and the supercell,
and it depends probably also on Cu ordering.
The estimation of $\lambda$ show that conventional superconductivity
based on electron-phonon coupling is probable, with a $T_C$ of the order of 
a few K. 
Despite the changes in electronic
structure, from a low DOS semi-metal in pure TiSe$_2$ to a high DOS 
material at high Cu doping, it is the Ti states that makes $\lambda$ large.
 Some results, like the large Stoner factor, indicate that disorder in 
the real material is important
to moderate the very large $N(E_F)$.
Effects of disorder will increase hybridization and can make
$\lambda$ larger even if the total DOS is lowered.
The particular situation with $E_F$ near a shoulder in the DOS,
and electron doping via Cu-impurities, turns out to give strong volume dependence
of $N(E_F)$ in addition to the dependence on $x$.
Further work is needed to sort out the precise effects of disorder, but it can already be 
predicted that not only variations of $x$, but also $P$-variations,
will be determining for
the properties of Cu$_x$TiSe$_2$.

\end{document}